\documentclass[a4paper]{aipproc}

\layoutstyle{6x9}

\def\be{\begin{equation}}
\def\ee{\end{equation}}
\def\bea{\begin{eqnarray}}
\def\eea{\end{eqnarray}}

\def\frac#1#2{{\textstyle{#1\over#2}}}
\def\darr#1{\raise1.5ex\hbox{$\leftrightarrow$}\mkern-16.5mu #1}
\def\){\right)}
\def\({\left( }
\def\]{\right] }
\def\[{\left[ }
\def\si{{}^1\kern-.14em S_0}
\def\siii{{}^3\kern-.14em S_1}
\def\diii{{}^3\kern-.14em D_1}

\def\cpt{\chi PT}

\def\vkay{{\vec k}}
\def\vkayprime{{{\vec k}\, '}}
\def\vpee{{\vec p}}
\def\vpeeprime{{{\vec p}\, '}}

\def\si{{}^1\kern-.14em S_0}
\def\siii{{}^3\kern-.14em S_1}
\def\diii{{}^3\kern-.14em D_1}

\def\lsim{\mathrel{\rlap{\lower4pt\hbox{\hskip1pt$\sim$}}        
    \raise1pt\hbox{$<$}}}         
\def\gsim{\mathrel {\rlap{\lower4pt\hbox{\hskip1pt$\sim$}}
    \raise1pt\hbox{$>$}}}         

\def\pone{{}^3\kern-.14em P_1}
\def\pzero{{}^3\kern-.14em P_0}
\def\ptwo{{}^3\kern-.14em P_2}

\begin{document}

\title 
      [$\chi$PT calcuations of electromagnetic deuteron reactions]
      {Probing the effectiveness: Chiral perturbation theory calculations 
of low-energy electromagnetic reactions on deuterium}


\author{Daniel R. Phillips}{
  address={Department of Physics and Astronomy, Ohio University, 
	Athens, OH 45701, U.~S.~A.;\\
Email: phillips@phy.ohiou.edu.}
}

\copyrightyear  {2001}

\begin{abstract}
I summarize three recent calculations of electromagnetic reactions on
deuterium in chiral perturbation theory. All of these calculations
were carried out to $O(Q^4)$, i.e. next-to-next-to-leading order.  The
reactions discussed here are: elastic electron-deuteron scattering,
Compton scattering on deuterium, and the photoproduction of neutral
pions from deuterium at threshold.
\end{abstract}

\date{\today}

\maketitle

\section{Introduction}

Effective field theory (EFT) is a technique commonly used in particle
physics to deal with problems involving widely-separated energy
scales. It facilitates the systematic separation of the effects of
high-energy physics from those of low-energy physics. In
strong-interaction physics the low-energy effective theory is chiral
perturbation theory ($\chi$PT)~\cite{Br95}. Here the low-energy
physics is that of nucleons and pions interacting with each other in a
way that respects the spontaneously-broken approximate chiral symmetry
of QCD. Higher-energy effects of QCD appear in $\chi$PT as
non-renormalizable contact operators.  The EFT yields amplitudes which
can be thought of as expansions in the ratio of nucleon or probe
momenta (denoted here by $p$ and $q$) and the pion mass to the scale
of chiral-symmetry breaking, $\Lambda_{\chi \rm{SB}}$, which is of
order the mass of the $\rho$ meson. The existence of a small parameter
$Q=p/\Lambda_{\chi \rm{SB}}$, $q/\Lambda_{\chi \rm{SB}}$,
$m_\pi/\Lambda_{\chi \rm{SB}}$, means that hadronic processes can be
computed in a controlled way.

The momentum scale of binding in light nuclei is of order $m_\pi$ and
so we should be able to calculate the response of such nuclei to
low-energy probes using $\chi$PT. The result is a
systematically-improvable, model-independent description.  Here I will
describe a few recent calculations using this approach. Section 2
outlines the $\chi$PT expansion, and sketches the implications of
$\chi$PT for calculations of the $NN$ interaction. Section 3 then
looks at electron-deuteron scattering in $\chi$PT as a probe of
deuteron structre. Section 4 examines the use of Compton scattering on
the deuteron as a way to extract neutron polaraizabilities.  I
conclude in Section 5 with a discussion of neutral pion
photoproduction on deuterium at threshold. In these last two reactions
I will argue that $\chi$PT's successful reproduction of the
experimental data is in no small part due to its consistent treatment
of the chiral structure of the nucleon and the deuteron. 

\section{Power counting and deuteron wave functions}

\subsection{Power counting}

Consider an elastic scattering process on the deuteron whose amplitude
we wish to compute. If $\hat{O}$ is the transition operator for this
process then the amplitude in question is simply $\langle \psi|
\hat{O} |\psi \rangle$, with $|\psi \rangle$ the deuteron wave
function. In this section, we follow Weinberg~\cite{We90,We91,We92}, and divide
the formulation of a systematic expansion for this amplitude into two
parts: the expansion for $\hat{O}$, and the construction of $|\psi
\rangle$.

Chiral perturbation theory gives a systematic expansion
for $\hat{O}$ of the form

\begin{equation} 
\hat{O}=\sum_{n=0}^\infty \hat{O}^{(n)},
\label{eq:expansion}
\end{equation} 
where we have labeled the contributions to $\hat{O}$ by their order
$n$ in the small parameter $Q$ defined above.
Eq.~(\ref{eq:expansion}) is an operator statement, and the nucleon
momentum operator $\hat{p}$ appears on the right-hand side.  However,
the only quantities which ultimately affect observables are
expectation values such as $\langle \psi| \hat{p} |\psi \rangle$. For
light nuclei this number is generically small compared to
$\Lambda_{\chi {\rm SB}}$.

To construct $\hat{O}^{(n)}$ one first writes down
the vertices appearing in the chiral Lagrangian up to order $n$. One
then draws all of the two-body, two-nucleon-irreducible, Feynman
graphs for the process of interest which are of chiral order $Q^n$. The
rules for calculating the chiral order of a particular graph are:

\begin{itemize}
\item Each nucleon propagator scales like $1/Q$;

\item Each loop contributes $Q^4$;

\item Graphs in which both particles participate in the reaction
acquire a factor of $Q^3$;

\item Each pion propagator scales like $1/Q^2$;

\item Each vertex from the $n$th-order piece of the chiral Lagrangian
contributes $Q^n$.  
\end{itemize}

In this way we see that more complicated graphs, involving two-body
mechanisms, and/or higher-order vertices, and/or more loops, are suppressed by
powers of $Q$.

\subsection{Deuteron wave functions}

There remains the problem of constructing a deuteron wave function
which is consistent with the operator $\hat{O}$. Weinberg's
proposal was to construct a $\chi$PT expansion in
Eq.~(\ref{eq:expansion}) for the $NN$ potential $V$, and then solve
the Schr\"odinger equation to find the deuteron (or other nuclear)
wave function~\cite{We90,We91,We92}.  Recent calculations have shown that the
$NN$ phase shifts can be understood, and deuteron bound-state static
properties reliably computed, with wave functions derived from
$\chi$PT in this way~\cite{Or96,KW97,Ep99,Re99,Ma01}. 

Now for $\chi$PT in the Goldstone-boson and single-nucleon sector loop
effects are generically suppressed by powers of the small parameter
$Q$. In zero and one-nucleon reactions the power counting in $Q$
applies to the amplitude, and not to the two-particle
potential. However, the existence of the deuteron tells us immediately
that a power counting in which loop effects are suppressed cannot be
correct for the two-nucleon case, since if it were there would be no
$NN$ bound state. Weinberg's proposal to instead power-count the
potential is one response to this dilemma. However, its consistency
has been vigorously debated in the literature (see~\cite{vK99,Be00}
for reviews).  Recently Beane {\it et al.}~\cite{Be01} have resolved
this discussion, by showing that Weinberg's proposal is consistent in
the $\siii-\diii$ channel.

One way to understand the $\chi$PT power-counting for deuteron wave
functions is to examine the deuteron wave function in three different
regions. Firstly, in the region $R \gg 1/m_\pi$ the deuteron wave
function is described solely by the asymptotic normalizations $A_S$,
$A_D$, and the binding energy $B$. These quantities are observables,
in the sense that they can be extracted from phase shifts by an
analytic continuation to the deuteron pole.

The second region corresponds to $R \sim 1/m_\pi$. Here pion exchanges
play a role in determining the $NN$ potential $V$, and, associatedly,
the deuteron wave functions $u$ and $w$. The leading effect comes from
iterated one-pion exchange---as has been known for at least fifty
years. Calculations with one-pion exchange (OPE) defining the
potential in this regime will be referred to below as
``leading-order'' (LO) calculations for the deuteron wave
function. Corrections at these distances come from two-pion exchange,
and these corrections can be consistently calculated in $\chi$PT.
They are suppressed by powers of the small parameter $Q$, and in fact
the ``leading'' two-pion exchange is suppressed by $Q^2$ relative to
OPE. This two-pion exchange can be calculated from vertices in ${\cal
L}_{\pi N}^{(1)}$ and its inclusion in the $NN$ potential results in
the so-called ``NLO'' calculation described in detail in
Ref.~\cite{Ep99}. Corrections to this two-pion-exchange result from
replacing one of the NLO two-pion-exchange vertices by a vertex from
${\cal L}_{\pi N}^{(2)}$. This results in an additional suppression
factor of $Q$, or an overall suppression of $Q^3$ relative to OPE, and
an ``NNLO'' chiral potential~\cite{Or96,KW97,Ep99,Re99}. More details
on this can be found in the contributions of Epelbaum and Timmermans to
these proceedings.

Finally, at short distances, $R \ll 1/m_\pi$ we cutoff the chiral one
and two-pion-exchange potentials and put in some short-distance
potential whose parameters are arranged so as to give the correct
deuteron asymptotic properties.

\section{Elastic electron scattering on deuterium}

One quantitative test of this picture of deuteron structure is
provided by elastic electron-deuteron scattering. We thus turn our
attention to the deuteron electromagnetic form factors $G_C$, $G_Q$,
and $G_M$. These are matrix elements of the deuteron current $J_\mu$,
with:
\begin{eqnarray}
G_C&=&\frac{1}{3 e (1 + \eta)} \left(\left \langle 1\left|J^0\right|1 \right \rangle + 
\left \langle 0\left|J^0\right|0 \right \rangle + \left \langle -1\left|J^0\right|-1 \right \rangle
\right),\\ 
G_Q&=&\frac{1}{M_d^2} \frac{1}{2 e \eta (1 + \eta)} \left(\left \langle 0\left|J^0\right|0 \right \rangle - \left \langle 1\left|J^0\right|1 \right \rangle\right)\\ 
G_M&=&\frac{1}{\sqrt{2 \eta} (1 + \eta)} \left \langle1\left|J^+\right|0\right \rangle 
\end{eqnarray}
where we have labeled these (non-relativistic) deuteron states by the
projection of the deuteron spin along the direction of the momentum
transfer ${\bf q}$ and $\eta \equiv |{\bf q}|^2/(4 M_d^2)$. $G_C$,
$G_Q$, and $G_M$ are related to the experimentally-measured $A$, $B$,
and $T_{20}$ in the usual way, with $T_{20}$ being primarily sensitive
to $G_Q/G_C$ and $B$ depending only to $G_M$. Here we will compare
calculations of the charge and quadrupole form factor with the recent
extractions of $G_C$ and $G_Q$ from data~\cite{Ab00B}.

\begin{figure}[htbp]
   \caption{The impulse-approximation contribution to $G_C$ and $G_Q$
	is shown on the left, while two meson-exchange current
	mechanisms which would contribute were the deuteron not an
	isoscalar target are depicted in the middle and on the right.}
	\includegraphics[height=.15\textheight,width=.25\textwidth]{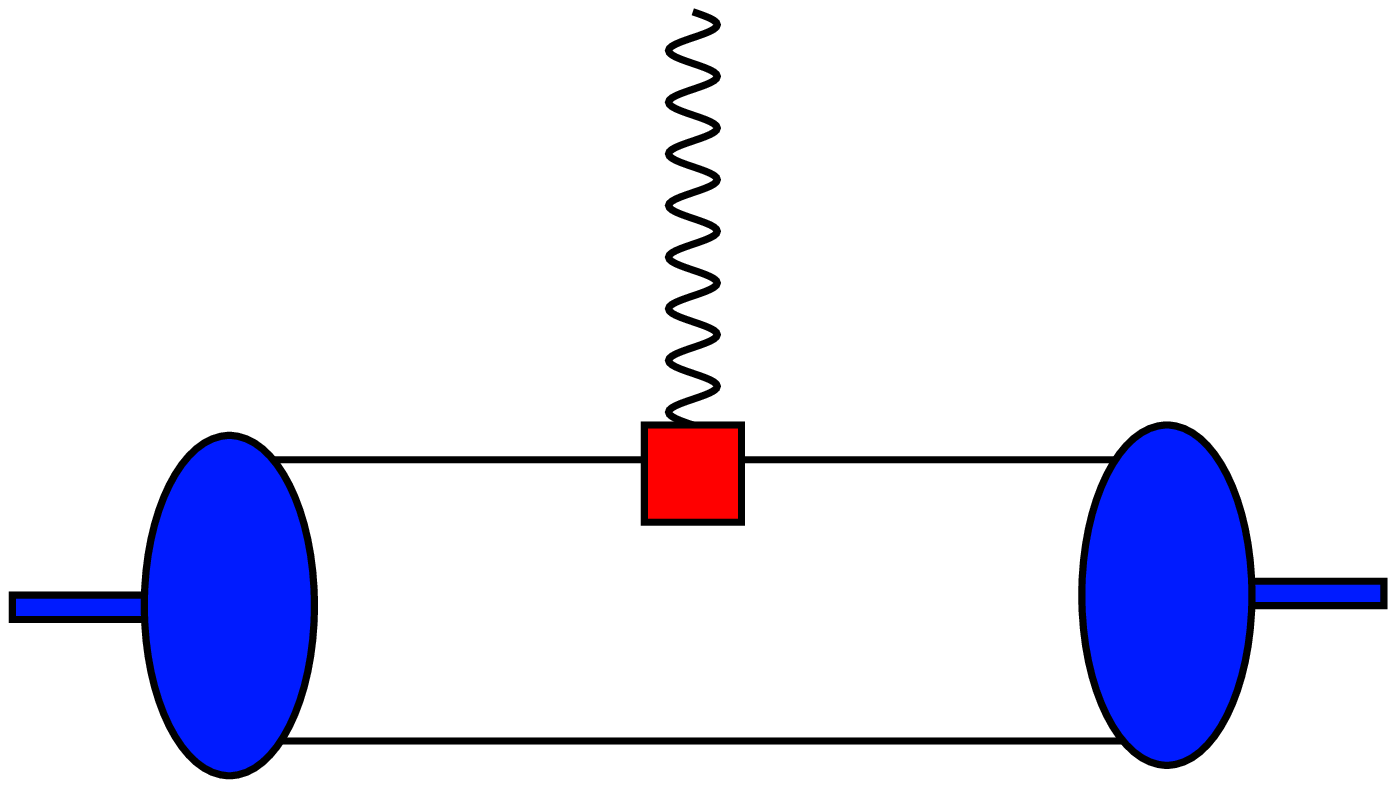}
	\includegraphics[height=.15\textheight,width=.6\textwidth]{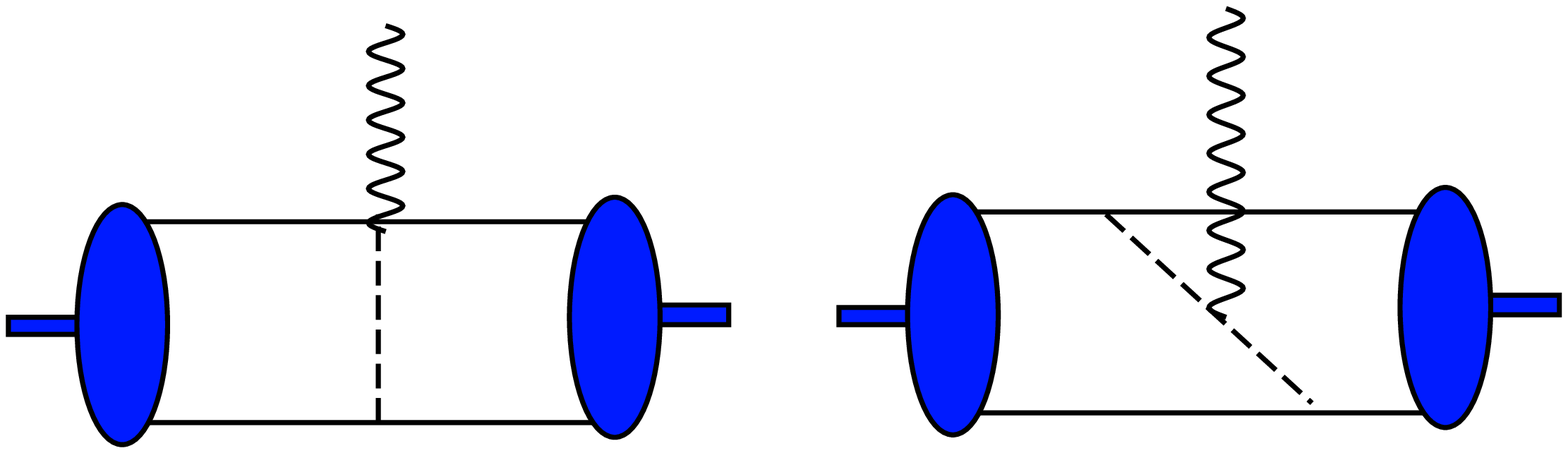}
\label{deuterongraphs}
\end{figure}

Both of these form factors involve the zeroth-component of the
deuteron four-current $J^0$. Here we split $J^0$ into two pieces: a
one-body part, and a two-body part. The one-body part of $J^0$ begins
at order $Q$ (where we are counting the proton charge $e \sim Q$) with
the impulse approximation diagram calculated with the non-relativistic
single-nucleon charge operator for strutcutreless
nucleons. Corrections to the single-nucleon charge operator from
relativistic effects and nucleon structure are suppressed by two
powers of $Q$, and thus arise at $O(Q^3)$, which is the
next-to-leading order (NLO) for $G_C$ and $G_Q$. At this order one
might also expect meson-exchange current (MEC) contributions, such as
those shown in Fig.~\ref{deuterongraphs}. However, all MECs
constructed with vertices from ${\cal L}_{\pi N}^{(1)}$ are isovector,
and so the first effect does not occur until N$^2$LO, or $O(Q^4)$,
where an $NN \pi \gamma$ vertex from ${\cal L}_{\pi N}^{(2)}$ replaces
the upper vertex in the middle graph of Fig.~\ref{deuterongraphs}, and
produces an isoscalar contribution to the deuteron charge operator. 
(This exchange-charge contribution was first derived by Riska~\cite{Ri84}.)

The most important correction that arises at NLO is the inclusion of
{\it nucleon} structure in $\chi$PT. At $O(Q^3)$ the isoscalar form
factors are dominated by short-distance physics, and so the only
correction to the point-like leading-order result comes from the
inclusion of the nucleon's electric radius, i.e.
\begin{equation}
{G_E^{(s)}}_{\mbox{$\chi$PT NLO}}=1 - \frac{1}{6} \langle r_E^{(s) \, 2}
\rangle q^2.
\end{equation}
This description of nucleon structure breaks down at momentum
transfers $q$ of order 300 MeV. There is a concomitant failure in the
description of eD scattering data~\cite{MW01,Ph01}.  Consequently, in
order to focus on {\it deuteron} structure, in the results presented
below I have chosen to circumvent this issue by using a ``factorized''
inclusion of nucleon structure~\cite{Ph01}. This facilitates the inclusion of
experimentally-measured single-nucleon form factors in the calculation,
thereby allowing us to test how far the theory is able to describe the
{\it two-body} dynamics that takes place in eD scattering.

\begin{figure}[thbp]
\caption{The deuteron charge and quadrupole form factors to order
$Q^4$ in chiral perturbation theory. The experimental data is taken
from the compilation of Ref.~\cite{Ab00B}. $G_Q$ is in units of
fm$^2$.}  \includegraphics[height=0.3\textheight]{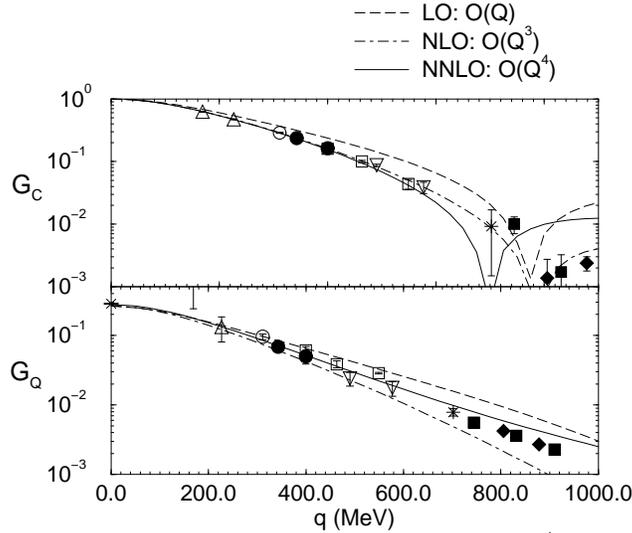}
\label{fig-FCFQ}
\end{figure}

The results for $G_C$ and $G_Q$ are shown in Fig.~\ref{fig-FCFQ}. The
figure demonstrates that convergence is quite good below $q \sim 700$
MeV---especially for $G_C$. The results shown are for the NLO chiral
wave function, but the use of the NNLO chiral wave function, or indeed
of simple wave functions which include only one-pion exchange, do not modify
the picture greatly below $q=700$ MeV~\cite{Ph01}. It is also clear
that---provided information from eN scattering is taken into
account---$\chi$PT is perfectly capable of describing the charge and
quadrupole form factors of deuterium at least as far as the minimum in
$G_C$. This result is extremely encouraging for the application of EFT
to light nuclei.

$G_M$ can be obtained in a similar way, but, importantly, the LO
contribution to $G_M$ is $O(Q^2)$. Furthermore, no two-body mechanism
enters until $O(Q^5)$, when an undetermined two-body counterterm
appears~\cite{MW01,Pa99}. Results for $F_M$ at $O(Q^4)$ turn out to be
of similar quality to those for $F_C$~\cite{MW01,Ph01}, but are
somewhat more sensitive to short-distance physics, as was expected
given the presence of the $O(Q^5)$ counterterm in this observable.

The static properties of the deuteron obtained in this expansion are
also generically quite reasonable, and have good convergence
properties. The one exception to this is the deuteron quadrupole
moment, $Q_d$, which, is underpredicted by about 4\%---as is also true
in all modern potential-model calculations~\cite{CS98}. However, such
an underprediction is not unexpected in $\chi$PT since simple
estimates of the effect on $Q_d$ of higher-order terms in the chiral
Lagrangian suggest that a discrepancy of order 5\% is to be expected
at $O(Q^4)$. $Q_d$ is rather sensitive to short-distance physics and
it transpires that higher-order counterterms have a larger effect on
it than on $G_C$~\cite{Ch99,PC99}. This way of understanding the
``$Q_d$ puzzle'' is one example of the way in which $\chi$PT can
assist in the analysis of electromagnetic currents for few-body
systems.  For an analogous application of $\chi$PT/EFT ideas to the
important solar reaction $pp \rightarrow d e^+ \nu_e$ see
Ref.~\cite{Pa01} and the contribution of Marcucci to these
proceedings.

\section{Compton scattering on deuterium}
\label{sec-gammagammad}

\noindent
Compton scattering on the nucleon at low energies is a fundamental
probe of the long-distance structure of these hadrons.  This process
has been studied in $\cpt$ in Ref.~\cite{Br92,Br92B}, where the
following results for the proton polarizabilities were obtained at LO:
\begin{eqnarray}
\alpha_p={{5 e^2 g_A^2}\over{384 \pi^2 f_\pi^2 m_\pi}} =12.2 \times
10^{-4} \, {\rm fm}^3; \qquad \beta_p={{e^2 g_A^2}\over{768 \pi^2
f_\pi^2 m_\pi}}= 1.2 \times 10^{-4} \, {\rm
fm}^3.
\end{eqnarray}
Recent experimental values for the proton polarizabilities are
\cite{To98}
\begin{eqnarray}
\alpha_p + \beta_p=13.23 \pm 0.86^{+0.20}_{-0.49} \times 10^{-4} \, {\rm fm}^3,
\nonumber\\
\alpha_p - \beta_p=10.11 \pm 1.74^{+1.22}_{-0.86} \times 10^{-4} \, {\rm fm}^3,
\label{polexpt}
\end{eqnarray}
where the first error is a combined statistical and systematic error,
and the second set of errors comes from the theoretical model
employed. These values are in good agreement with the $\cpt$
predictions.

Chiral perturbation theory also predicts $\alpha_n=\alpha_p$,
$\beta_n=\beta_p$ at this order.  The neutron polarizabilities
$\alpha_n$ and $\beta_n$ are difficult to measure, due to the absence
of suitable neutron targets, and so this prediction is not well
tested.  One way to extract $\alpha_n$ and $\beta_n$ is to perform
Compton scattering on nuclear targets. Coherent Compton scattering on
a deuteron target has been measured at $E_\gamma=$ 49 and 69 MeV by
the Illinois group \cite{Lu94} and $E_\gamma= 84.2-104.5$ MeV at
Saskatoon~\cite{Ho00}.  The amplitude for Compton scattering on the
deuteron clearly involves mechanisms other than Compton scattering on
the individual constituent nucleons. Hence, the desire to extract
neutron polarizabilities argues for a theoretical calculation of
Compton scattering on the deuteron that is under control in the sense
that it accounts for {\it all} mechanisms to a given order in $\cpt$.

\begin{figure}[htbp]
   \caption{Graphs
   which contribute to Compton scattering on the deuteron at
   ${\cal O}(Q^2)$ (a) and ${\cal O}(Q^3)$ (b-d).  The
   sliced and diced blobs are from ${\cal L}_{\pi N}^{(3)}$ (c) and
   ${\cal L}_{\pi \gamma}^{(4)}$ (d).  Crossed graphs are not
   shown.}
\includegraphics[height=.25\textheight]{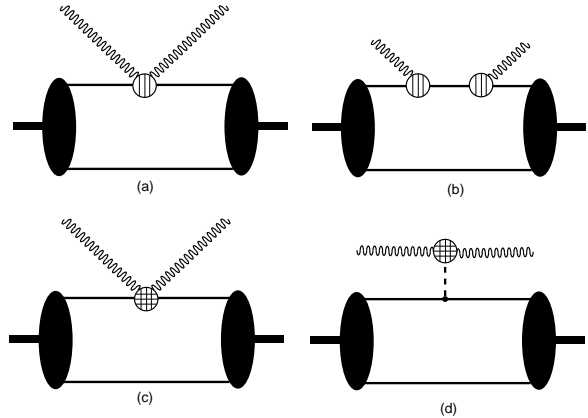}
\label{procomptonebodmod} 
\end{figure}

\begin{figure}[t,h.b,p]
\includegraphics[height=.15\textheight]{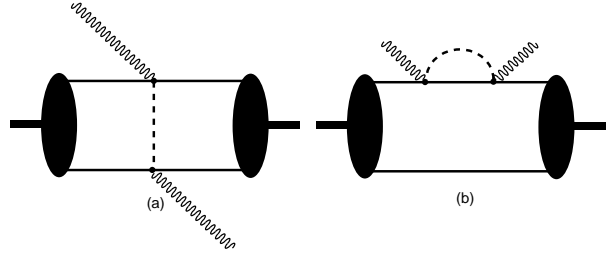}
   \caption{Graphs which contribute to Compton scattering 
on the deuteron at ${\cal O}(Q^3)$. 
Crossed graphs are not shown.}
\label{combined} 
\end{figure}

The Compton amplitude we wish to evaluate is (in the $\gamma d$
center-of-mass frame):
\begin{eqnarray}
T^{\gamma d}_{M' \lambda' M \lambda}(\vkayprime,\vkay)&=& \int
\frac{d^3p}{(2 \pi)^3} \, \, \psi_{M'}\left( \vpee + \frac{\vkay -
\vkayprime}{2}\right) \, \, T_{\gamma N_{\lambda'
\lambda}}(\vkayprime,\vkay) \, \, \psi_M(\vpee)\nonumber\\ 
&+& \int \frac{d^3p \, \, d^3p'}{(2 \pi)^6} \, \, \psi_{M'}(\vpeeprime) \, \,
T^{2N}_{\gamma NN_{\lambda' \lambda}}(\vkayprime,\vkay) \, \, \psi_M(\vpee)
\label{eq:gammad}
\end{eqnarray}
where $M$ ($M'$) is the initial (final) deuteron spin state, and
$\lambda$ ($\lambda'$) is the initial (final) photon polarization
state, and $\vkay$ ($\vkayprime$) the initial (final) photon
three-momentum, which are constrained to
$|\vkay|=|\vkayprime|=\omega$.  The amplitude $T_{\gamma N}$
represents the graphs of Fig.~\ref{procomptonebodmod} and
Fig.~\ref{combined}b where the photon interacts with only one nucleon.
The amplitude $T^{2N}_{\gamma NN}$ represents the graphs of
Fig.~\ref{combined}a where there is an exchanged pion between the two
nucleons.

The LO contribution to Compton scattering on the deuteron is shown in
Fig.~\ref{procomptonebodmod}(a).  This graph involves a vertex from
${\cal L}_{\pi N}^{(2)}$ and so is $O({Q^{2}})$.  This contribution is
simply the Thomson term for scattering on the proton. There is thus no
sensitivity to either two-body contributions {\it or} nucleon
polarizabilities at this order.  At $O({Q^{3}})$ there are several
more graphs with a spectator nucleon
(Figs.~\ref{procomptonebodmod}(b),(c),(d)), as well as graphs
involving an exchanged pion with leading order vertices
(Fig.~\ref{combined}(a)) and one loop graphs with a spectator nucleon
(Fig.~\ref{combined}(b))~\cite{Be99}. Graphs such as
Fig.~\ref{combined}(b) contain the physics of the proton and neutron
polarizabilities at $O(Q^3)$ in $\chi$PT.

\begin{figure}[!ht]
\includegraphics[height=.2\textheight,width=.3\textwidth]{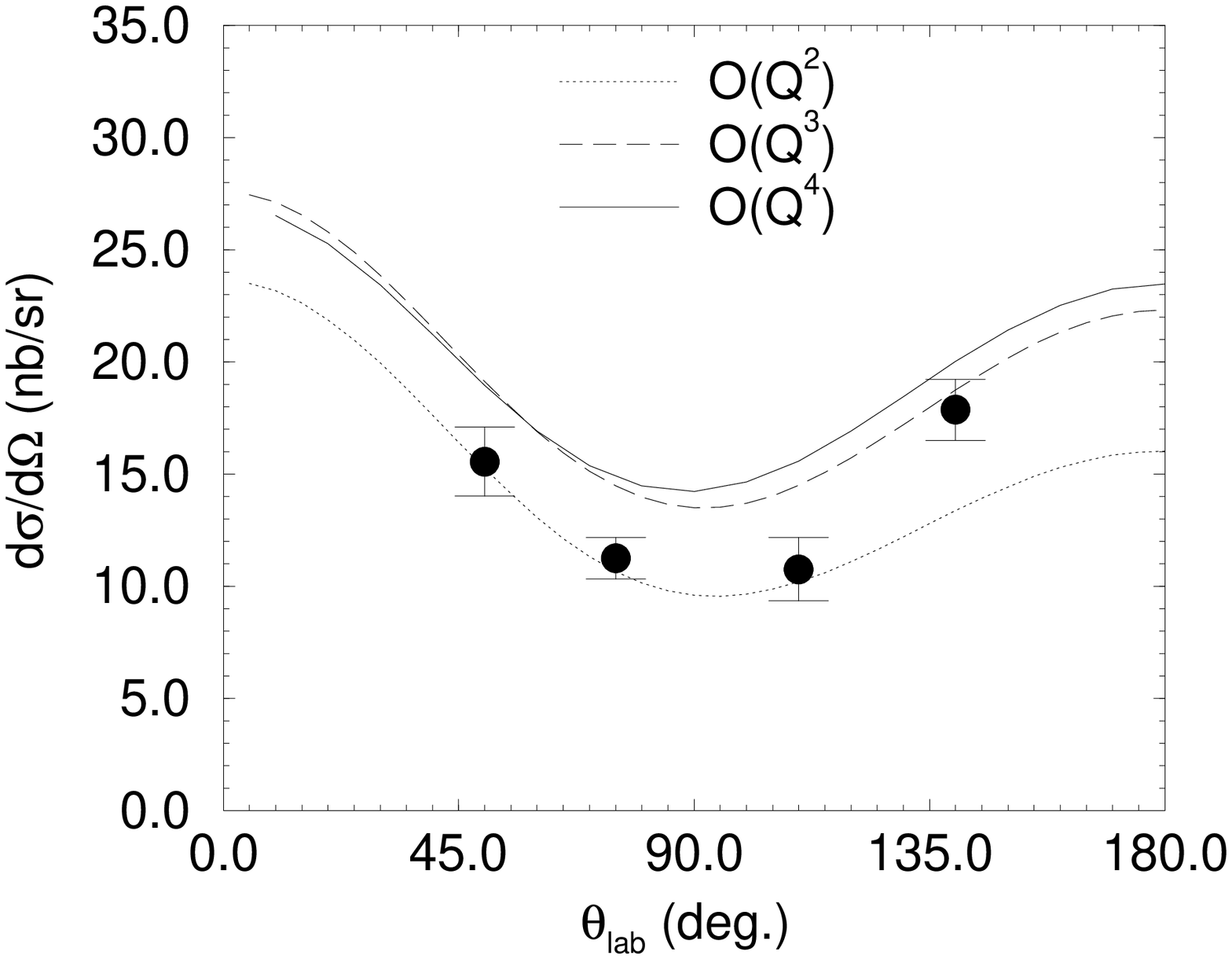}
\includegraphics[height=.2\textheight,width=.3\textwidth]{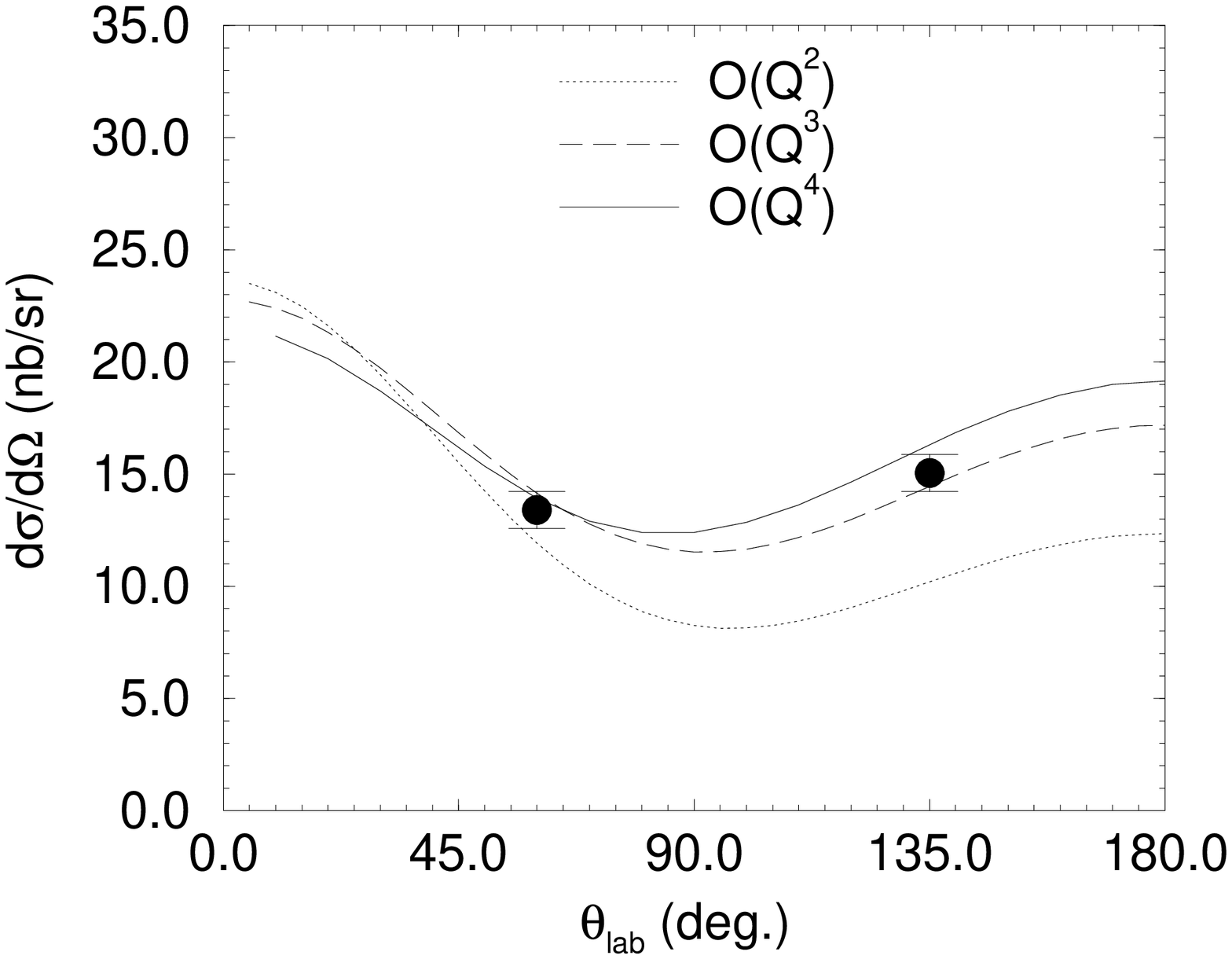}
\includegraphics[height=.2\textheight,width=.3\textwidth]{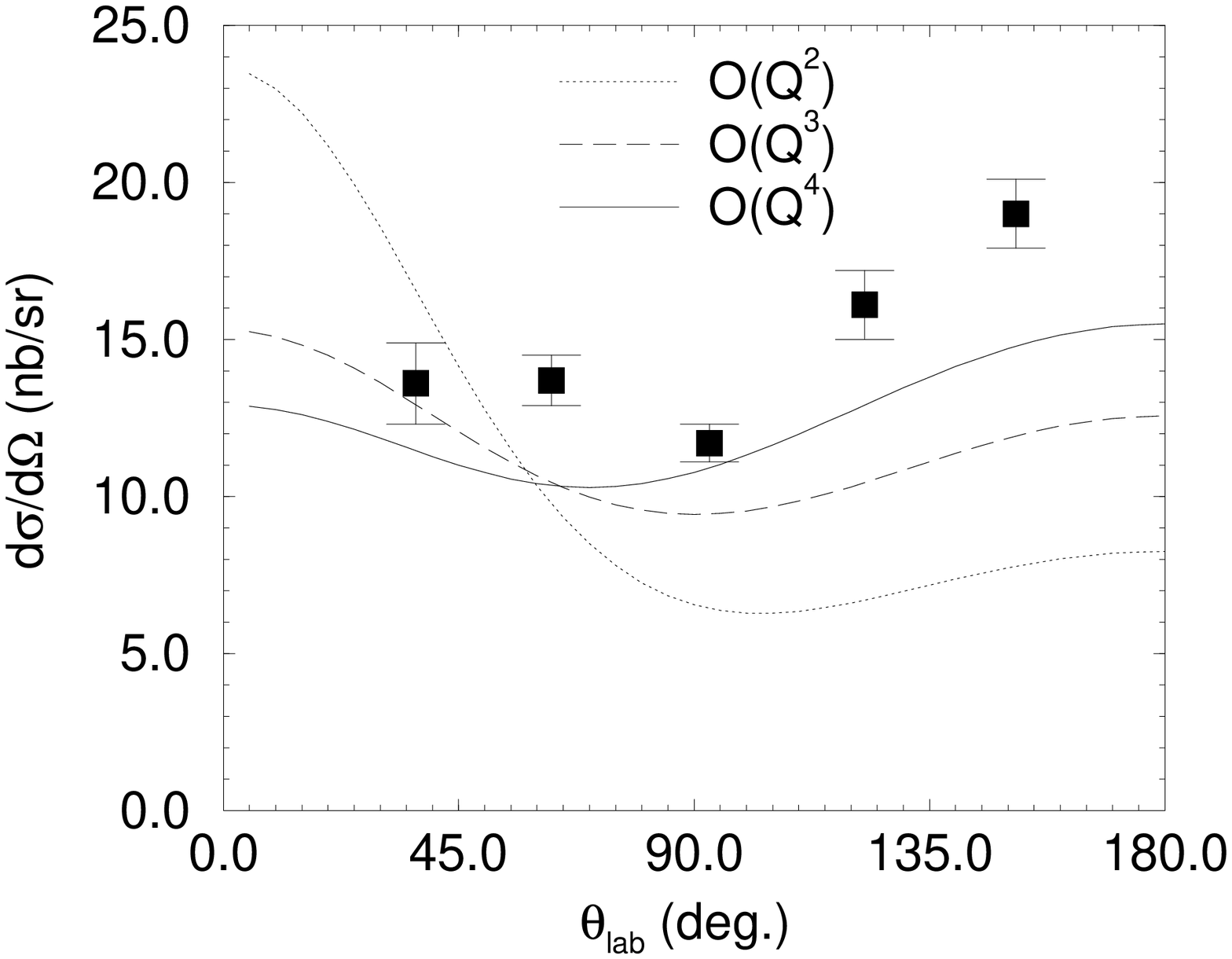}
\caption{Results of the $O(Q^2)$ (dotted line), $O(Q^3)$ (dashed
line), and $O(Q^4)$ (solid line) calculations for $E_\gamma=49~{\rm
MeV}$, $69~{\rm MeV}$ and $95~{\rm MeV}$ respectively from left to
right.}
\label{cvgplot}
\end{figure}

We employed a variety of wave functions $\psi$, and found only
moderate wave-function sensitivity. Results shown here are generated
with the NLO chiral wave function of Ref.~\cite{Ep99}.
Fig.~\ref{cvgplot} shows the results at $E_\gamma = $ 49, 69, and 95
MeV.  For comparison we have included the calculation at $O(Q^2)$ in
the kernel, where the second contribution in Eq.~(\ref{eq:gammad}) is
zero, and the single-scattering contribution is given solely by
Fig.~\ref{procomptonebodmod}(a).  At $O(Q^3)$ all contributions to
the kernel are fixed in terms of known pion and nucleon parameters, so
to this order $\cpt$ makes {\it predictions} for deuteron Compton
scattering.  We also show the $O(Q^4)$ result which will be discussed
below. The curves indicate that higher-order corrections get larger as
$\omega$ is increased---as expected.

We have also shown the six Illinois data points at 49 and 69
MeV~\cite{Lu94} and the Saskatoon data at 95
MeV~\cite{Ho00}. Statistical and systematic errors have been added in
quadrature. It is quite remarkable how well the $O(Q^2)$ calculation
reproduces the 49 MeV data. However, the agreement is somewhat
fortuitous: there are significant $O(Q^3)$ corrections.  Note that at
these lower photon energies Weinberg power counting begins to break
down, since it is designed for $\omega \sim m_\pi$, and does {\it not}
recover the deuteron Thomson amplitude as $\omega \rightarrow 0$.
Correcting the power counting to remedy this difficulty appears to
improve the description of the 49 MeV data, without significantly
modifying the higher-energy results~\cite{Be99}. Meanwhile, the
agreement of the $O(Q^3)$ calculation with the 69 MeV data is very
good, although only limited conclusions can be drawn. These results
are not very different from other, potential-model,
calculations~\cite{Wl95,LL98,KM99}. They are also quite similar to
those obtained in $NN$ EFTs without explicit pions (see
Ref.~\cite{GR00}, and the contribution of Grie\ss hammer to these
proceedings).  However our calculation is the only one that does not
employ the polarizability approximation for the $\gamma$N amplitude.

At $O(Q^4)$ single-nucleon counterterms which shift the
polarizabilities enter the calculation. However, there are still no
two-body counterterms contributing to $\gamma d \rightarrow \gamma d$
at this order. In this sense Compton scattering on deuterium at
$O(Q^4)$ is analogous to the reaction $\gamma d \rightarrow \pi^0 d$
discussed below: an $O(Q^4)$ calculation allows us to test the
single-nucleon physics which is used to predict the results of
coherent scattering on deuterium, since there are no undetermined
parameters in the two-body mechanisms that enter to this order in the
chiral expansion.

The $O(Q^4)$ plot shown above is a partial calculation at that order.
It includes all two-body mechanisms at $O(Q^4)$, and some one-body
mechanisms, the latter being calculated in ways motivated by
dispersion-relation analyses. The values of the proton
polarizabilities used in Fig.~\ref{cvgplot} were taken from
Eq.~(\ref{polexpt}). Meanwhile the (disputed) neutron-atom scattering
value of Ref.~\cite{Sc91} was employed for $\alpha_n$ and the Baldin
sum rule used to fix $\beta_n$. To demonstrate the sensitivity to
$\alpha_n$, we can also use the freedom in the single-nucleon
amplitude at $O(Q^4)$ to fit the SAL data using the incomplete
$O(Q^4)$ calculation.  A reasonable fit at backward angles can be
achieved with $\alpha_n =4.4$ and $\beta_n =10$. These numbers are in
startling disagreement with the $O(Q^3)$ $\cpt$ expectations. We have
also plotted the cross-section at 69 MeV with $\alpha_n =4.4$ and
$\beta_n =10$: this curve misses the Illinois data. (Similar results
were found in a potential model in Ref.~\cite{LL98}.)  This situation
poses an interesting theoretical puzzle. A full $O(Q^4)$ calculation
in $\cpt$ using the recently derived single-nucleon Compton
amplitude~\cite{McG01} is necessary before firm conclusions can be
drawn. Such a calculation is in progress~\cite{Be02}.

\begin{figure}[htbp]
\includegraphics[height=.2\textheight,width=.35\textwidth]{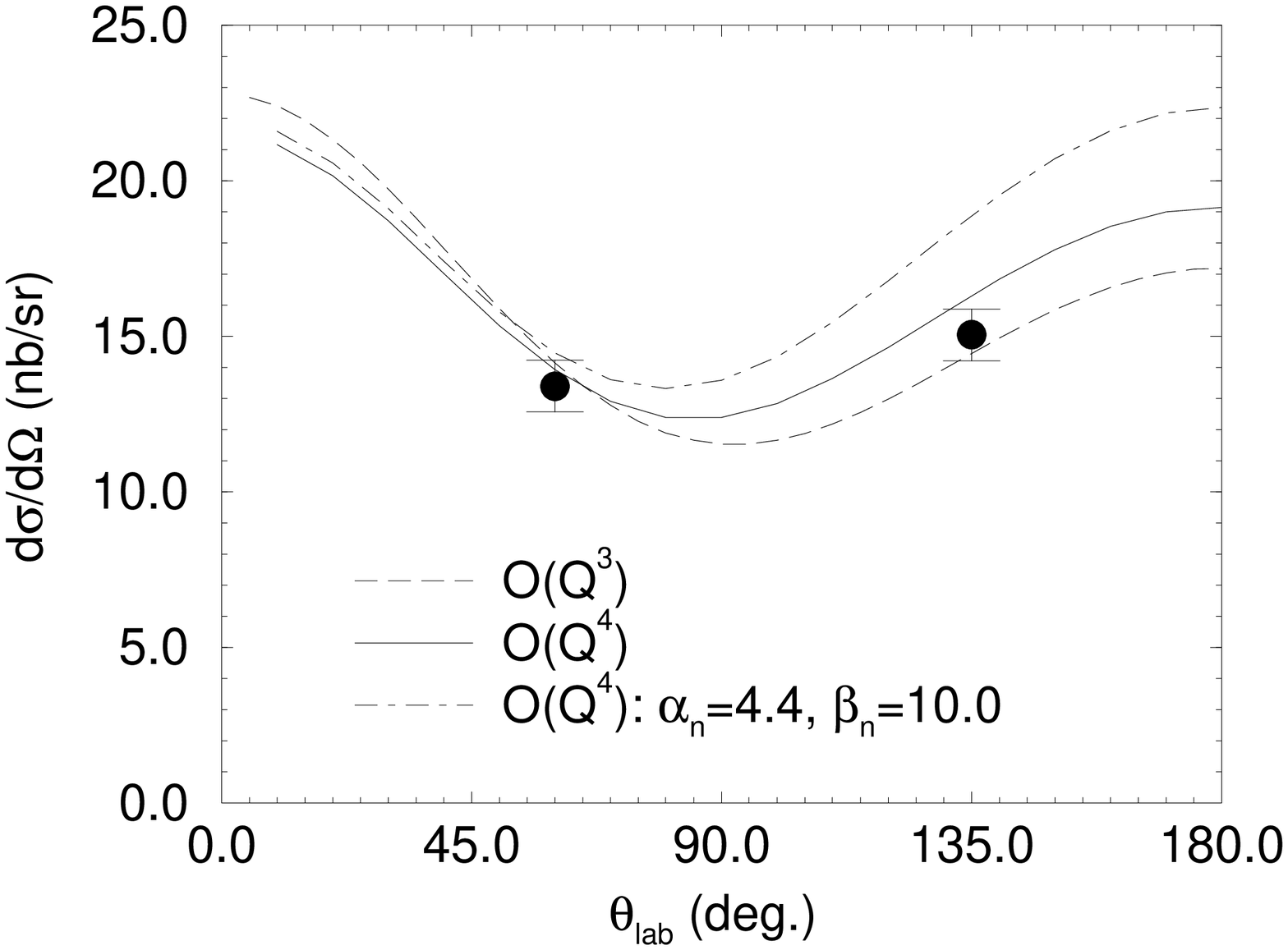}
\includegraphics[height=.2\textheight,width=.35\textwidth]{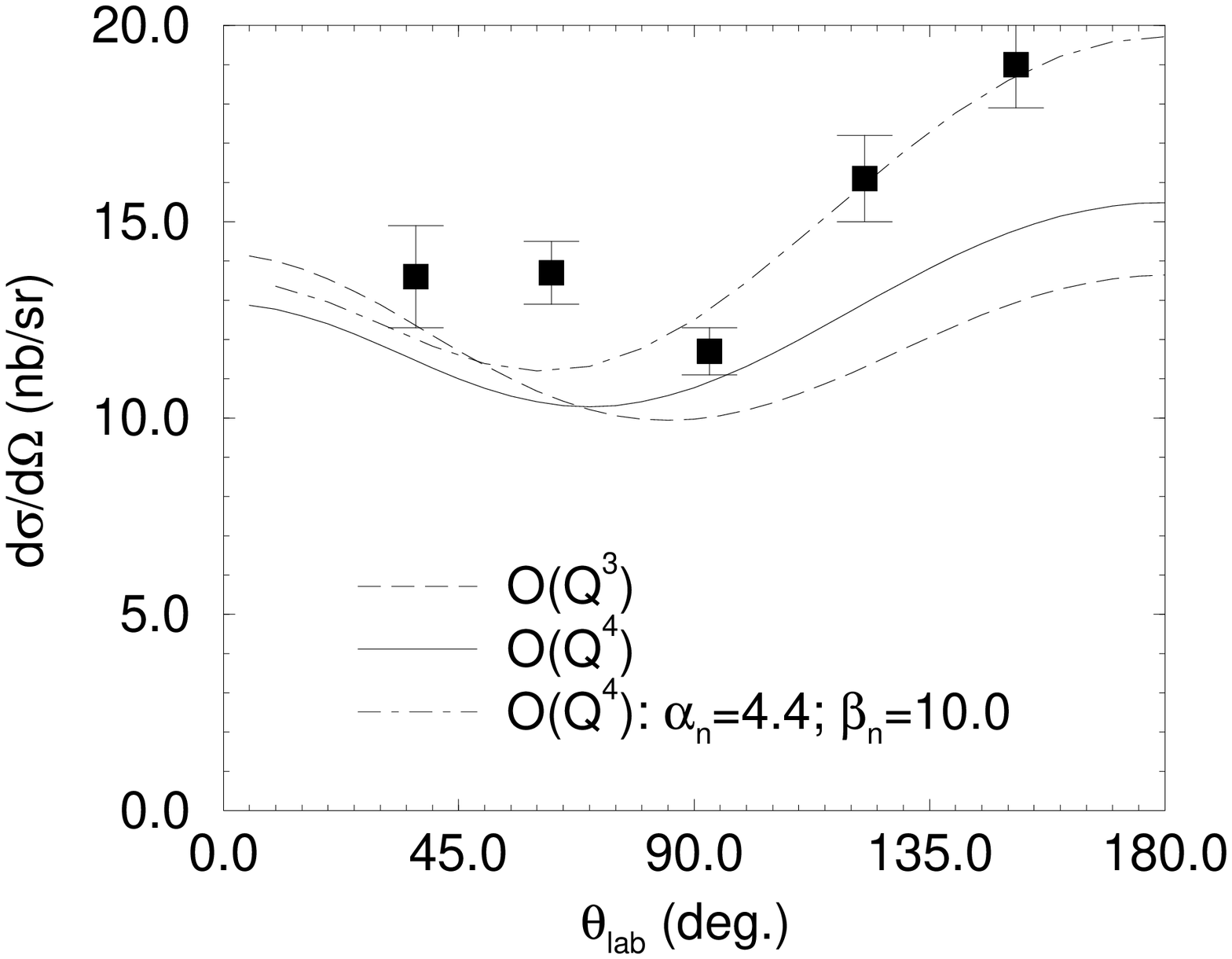}
\caption{Results of $O(Q^3)$ (dashed line), partial $O(Q^4)$ (solid
line), and partial $O(Q^4)$ with modified $\alpha_n$ (dot-dashed
line), $\chi$PT calculations for $E_\gamma=69~{\rm MeV}$ (left panel)
and $95~{\rm MeV}$ (right panel).}
\label{SALplot}
\end{figure}

\section{Neutral pion photoproduction on deuterium}
\label{sec-gammadpi0d}

\noindent
Pion photoproduction on the nucleon near threshold has been studied up
to $O(Q^4)$ in $\cpt$ with the delta integrated out in
Ref.~\cite{Br96C}.  The differential cross-section at threshold is
given solely by $E_{0+}$, the electric dipole amplitude.  In $\chi$PT
neutral pion photoproduction on the nucleon does not begin until
$O(Q^2)$, where there is a tree-level contribution from a $\gamma \pi
NN$ vertex. Then at $O(Q^3)$ there are tree-level contributions
involving the magnetic moment of the nucleon. The sum of these
$O(Q^2)$ and $O(Q^3)$ effects reproduces an old ``low-energy theorem''.
Also at $O(Q^3)$ there occur finite loop corrections where the photon
interacts with a virtual pion which then rescatters on the nucleon.
This large quantum effect was missed in the old ``low-energy theorem''
and is absent in most models. At $O(Q^4)$ there are loops with
relativistic corrections, together with a counterterm.  The proton
amplitude that results from fitting this counterterm to data produces
energy-dependence in relatively good agreement with the recent results
from Mainz and Saskatoon.  At the same order there is also a
prediction for the near-threshold behavior of the $\gamma n\rightarrow
\pi^0 n$ reaction; the cross-section is considerably larger than that
obtained in models that omit the important pion-cloud $O(Q^3)$
diagrams. In fact, the $O(Q^4)$ prediction is that the cross section
for $\gamma n \rightarrow \pi^0 n$ is about four times larger than
that for neutral pion photoproduction on the proton. A good way to
test this prediction is to study neutral pion photoproduction on the
deuteron.

If deuterium is to be used as a target in this particular process then
two-body mechanisms that contribute to the reaction must be
calculated. $\chi$PT provides an ideal way to do this, as was
demonstrated by Beane and collaborators~\cite{Be97}.

Consider the reaction $\gamma (k) + d(p_1) \to
\pi^0 (q) + d(p_2)$ in the threshold region, $\vec{q} \simeq 0$, where
the pion is in an $S$ wave with respect to the center-of-mass (cm)
frame.  For real photons the threshold differential
cross section is:
\begin{equation}
{{|{\vec k}|}\over{|{\vec q\,}|}}\;{{d\sigma}\over{d\Omega}}
\bigg|_{|{\vec q\,}|=0} =\frac{8}{3}{E_d^2}.
\end{equation}
We now present the results of the chiral expansion of the dipole
amplitude $E_d$ to $O(Q^4)$. 

Two-body contributions to $E_d$ do not begin until $O(Q^3)$. Thus, to
$O(Q^4)$ we have
\begin{eqnarray}
E_d&=&\langle \psi|\hat{O}_{ob}|\psi \rangle + \langle
\psi|\hat{O}_{tb}^{(3)}|\psi \rangle + + \langle
\psi|\hat{O}_{tb}^{(4)}|\psi \rangle \\ &\equiv& {E^{ss}_d} +
E_d^{tb,3} + E_d^{tb,4},
\end{eqnarray}
where we have explicitly isolated the $O(Q^3)$ and $O(Q^4)$ two-body
contributions to the operator $\hat{O}$, and $|\psi \rangle$ is a deuteron
wave function.

The single-scattering contribution to $E_d$, $E_d^{ss}$, is given by
all diagrams where the photon is absorbed and the pion emitted from
one nucleon with the second nucleon acting as a spectator, i. e. the
impulse approximation result.  Since the $\chi$PT results for the
elementary $S$-wave pion production amplitudes to $O(Q^4)$
\cite{Br96C} are known there is an $O(Q^4)$ $\chi$PT {\it prediction}
for $E^{ss}_d$. This is the prediction we want to test.

\begin{figure}[t,h.b,p]
\includegraphics[height=.15\textheight]{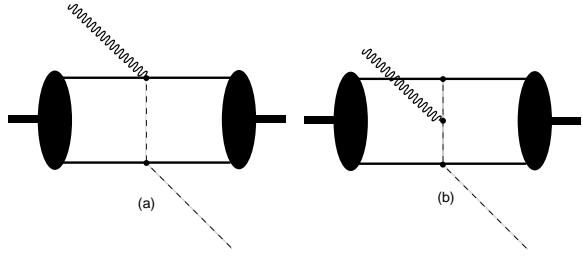}
\caption{
Two-nucleon graphs which contribute to neutral pion photoproduction
at threshold to $O(Q^3)$ (in the Coulomb gauge).
All vertices come from ${\cal L}^{(1)}_{\pi N}$.}
\label{q3newmod}
\end{figure}

But, before comparing this prediction with the experimental data we
must compute the two-body contributions to $E_d$. Those of $O(Q^3)$
which survive at threshold, in the Coulomb gauge, are shown in
Fig.~\ref{q3newmod}~\cite{Be95}.

\begin{figure}[t,h.b,p]
\includegraphics[height=.15\textheight]{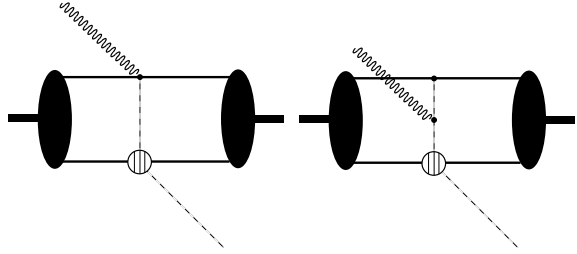}
\caption{
Characteristic two-nucleon graphs contributing at $O(Q^4)$ to neutral pion photoproduction.
The hatched circles denote an insertion from ${\cal L}_{\pi N}^{(2)}$.}
\label{fig3mod2}
\end{figure}

At $O(Q^4)$, we have to consider the two-nucleon diagrams---some of
which are shown in Fig. \ref{fig3mod2}, with the blob characterizing
an insertion from ${\cal L}_{\pi N}^{(2)}$. There are also
relativistic corrections to the graphs in Fig.~\ref{q3newmod}.

One can show that the only terms that survive at threshold result in
insertions $\sim 1/2M$, $\sim g_A/2M$ and $\sim \kappa_{0,1}$.  To
$O(Q^4)$ there are no four-nucleon operators contributing to the
deuteron electric dipole amplitude, and so {\it no new, undetermined
parameters appear}. The only free parameter is fixed in $\gamma p
\rightarrow \pi^0 p$, and so a genuine {\it prediction} can be, and
was, made for the reaction $\gamma d \rightarrow \pi^0 d$ at
threshold.

We now present results. Using a variety of deuteron wave functions the
single-scattering contribution is found to be~\cite{Be97}:
\begin{equation} 
\label{Edssval}
E^{ss}_{d}=(0.36\pm 0.05) \times 10^{-3}/m_{\pi^+}.
\end{equation}
The sensitivity of the single-scattering contribution $E_d^{ss}$ to the
elementary neutron amplitude may be parameterized by:
\begin{equation}
E_d^{ss} = \left[ 0.36 - 0.38 \cdot (2.13 - E_{0+}^{{\pi ^0}n})
\right] \times 10^{-3}/m_{\pi^+}.
\label{sensi}
\end{equation}
The two-nucleon contribution is evaluated at $O(Q^3)$ using different
deuteron wave functions~\cite{Be97}. The results prove to be largely
insensitive to the choice of wave function, as do the two-body
contributions at $O(Q^4)$. Choosing the AV18 wave function for
definiteness the results are summarized in Table
\ref{Edconts}\footnote{Similar answers were obtained with the wave
functions of Ref.~\cite{Or96}, which should be equivalent to the NNLO
wave function discussed above, up to higher-order terms.}. The
$O(Q^4)$ contributions give corrections of order 15\% to the $O(Q^3)$
two-nucleon terms.  We also observe that $E_d^{tb,4}$ is of the same
size as $E_d^{ss}$, clearly demonstrating the need to go to this order
in the expansion. This gives us confidence that the $\chi$PT expansion
is controlled, and that we may compare experimental data with the
$O(Q^4)$ prediction~\cite{Be97} :
\begin{equation}
E_d = (-1.8 \pm 0.2)  \times 10^{-3}/m_{\pi^+}.
\label{Edtot}
\end{equation}

\begin{table}[t,h.b,p]
\caption{Values for $E_{d}$ in units of $10^{-3}/m_{\pi^+}$
from one-nucleon contributions ($1N$) up to $O(Q^4)$,
two-nucleon kernel ($2N$) at $O(Q^3)$ 
and at $O(Q^3)$ ,
and their sum ($1N+2N$).}
\label{Edconts}
\vspace{0.2cm}
\begin{tabular}{|cccc|}
\hline
$1N$   & \multicolumn{2}{c}{$2N$} &    $1N+2N$        \\
\cline{2-3}
$Q+Q^2+Q^3+Q^4$ & $Q^3$ & $Q^4$   
                                          & $Q+Q^2+Q^3+Q^4$  \\
\hline
 0.36  & $-$1.90 & $-$0.25                  & $-$1.79 \\
\hline
\end{tabular}
\end{table}

\begin{figure}[t,h.b,p]
\includegraphics[height=.27\textheight]{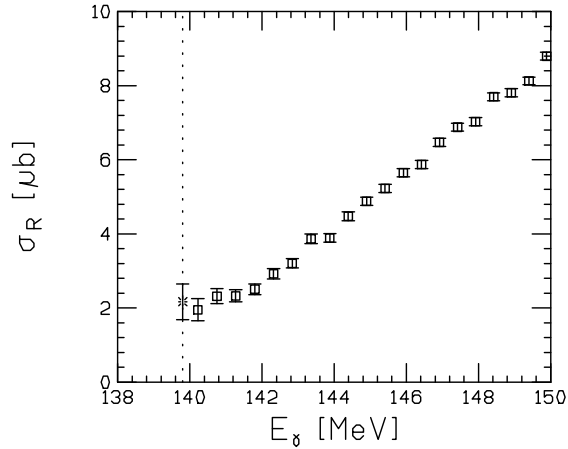}
\caption{
Reduced cross-section  $\sigma_R=(k/q)\sigma$ in $\mu$b
for neutral pion photoproduction as function of the photon 
energy in MeV.
Threshold is marked by a dotted line.
Squares are data points from Ref. \protect\cite{Be98}
and the star is the 
$\cpt$ {\it pre}diction of Ref. \protect\cite{Be97}.
Figure courtesy of Ulf Mei{\ss}ner.}
\label{piphotodata}
\end{figure}

To see the sensitivity to the elementary neutron amplitude, we set the
latter to zero and find $E_d = -2.6\times 10^{-3}/m_{\pi^+}$ (for the
AV18 potential). Thus $\chi$PT makes a prediction that differs
markedly from conventional models, with that difference arising
predominantly from a different result for $E_{0+}^{\pi^0 n}$.  An
experimental test of this prediction was carried out recently at
Saskatoon \cite{Be98}.  The results for the pion photoproduction
cross-section near threshold are shown in Fig.~\ref{piphotodata},
together with the $\cpt$ prediction at threshold. The agreement with
$\cpt$ to order $O(Q^4)$ is not better than a reasonable estimate of
higher-order terms, but $\chi$PT is clearly superior to models.  This is
compelling evidence of the importance of chiral loops, and also testimony
to the consistency and usefulness of $\chi$PT in analyzing low-energy
reactions on deuterium.

\begin{theacknowledgments}
I thank the organizers of ``Mesons and Light Nuclei'' for an enjoyable
meeting. I am also grateful to Silas Beane, Manuel Malheiro, and Bira
van Kolck for a profitable and enjoyable collaboration on $\gamma d
\rightarrow \gamma d$. Thanks are due to Silas for discussions on
$\gamma d \rightarrow \pi^0 d$ and numerous other subjects in
effective field theory. Finally, I am grateful to the U.~S. Insitute
for Nuclear Theory for its hospitality during the writing of this
paper.
\end{theacknowledgments}


\begin{thebibliography}{31}
\expandafter\ifx\csname natexlab\endcsname\relax\def\natexlab#1{#1}\fi
\providecommand{\enquote}[1]{``#1''}
\expandafter\ifx\csname url\endcsname\relax
  \def\url#1{\texttt{#1}}\fi
\expandafter\ifx\csname urlprefix\endcsname\relax\def\urlprefix{URL }\fi

\bibitem[Bernard et~al.(1995)]{Br95}
Bernard, V., Kaiser, N., and Mei\ss ner, U.-G., \emph{Int. Jour. of Mod. Phys.
  E}, \textbf{4}, 193 (1995).

\bibitem[Weinberg(1990)]{We90}
Weinberg, S., \emph{Phys. Lett.}, \textbf{B251}, 288 (1990).

\bibitem[Weinberg(1991)]{We91}
Weinberg, S., \emph{Nucl. Phys.}, \textbf{B363}, 3 (1991).

\bibitem[Weinberg(1992)]{We92}
Weinberg, S., \emph{Phys. Lett.}, \textbf{B295}, 114 (1992).

\bibitem[Ordon\'ez et~al.(1996)]{Or96}
Ordon\'ez, C., Ray, L., and van Kolck, U., \emph{Phys. Rev. C}, \textbf{53},
  2086 (1996).

\bibitem[Kaiser et~al.(1997)]{KW97}
Kaiser, N., Brockmann, R., and Weise, W., \emph{Nucl. Phys.}, \textbf{A625},
  758 (1997).

\bibitem[Epelbaum et~al.(1999)]{Ep99}
Epelbaum, E., Glockle, W., and Mei\ss ner, U.-G., \emph{Nucl. Phys.},
  \textbf{A671}, 295 (1999).

\bibitem[Rentmeester et~al.(1999)]{Re99}
Rentmeester, M.~C.~M., Timmermans, R.~G.~E., Friar, J.~L., and de~Swart, J.~J.,
  \emph{Phys. Rev. Lett.}, \textbf{82}, 4992 (1999).

\bibitem[Entem and Machleidt(2001)]{Ma01}
Entem, D.~R., and Machleidt, R., nucl-th/0107057.

\bibitem[van Kolck(1999)]{vK99}
van Kolck, U., \emph{Prog. Part. Nucl. Phys.}, \textbf{43}, 409 (1999).

\bibitem[Beane et~al.(2000)]{Be00} Beane, S.~R., Bedaque, P.~F.,
Haxton, W., Phillips, D.~R., and Savage, M.~J., in \emph{At the
frontier of particle physics---Handbook of QCD}, M. Shifman, ed. (World
Scientific, Singapore, 2000).

\bibitem[Beane et~al.(2001)]{Be01}
Beane, S.~R., Bedaque, P.~F., Savage, M.~J., and van Kolck, U.,
nucl-th/0104030.

\bibitem[Abbott et~al.(2000)]{Ab00B}
Abbott, D., et~al. [JLAB t20 Collaboration] \emph{Eur. Phys. J.},
\textbf{A7}, 421 (2000).

\bibitem[Riska (1984)]{Ri84}
Riska, D.~O., \emph{Prog.~Part.~Nucl.~Phys.}, \textbf{11}, 199 (1984).

\bibitem[Mei\ss ner and Walzl(2001)]{MW01}
Mei\ss ner, U.-G., and Walzl, M., \emph{Phys. Lett.}, \textbf{B513}, 37 (2001).

\bibitem[Phillips(2001)]{Ph01}
Phillips, D.~R., in preparation.

\bibitem[Park et~al.(2000)]{Pa99}
Park, T.-S., Kubodera, K., Min, D.-P., and Rho, M., \emph{Phys. Lett.},
  \textbf{B472}, 232 (2000).

\bibitem[Carlson and Schiavilla(1998)]{CS98}
Carlson, J., and Schiavilla, R., \emph{Rev. Mod. Phys.}, \textbf{70}, 743
  (1998).

\bibitem[Chen et~al.(1999)]{Ch99}
Chen, J.-W., Rupak, G., and Savage, M., \emph{Nucl. Phys.}, \textbf{A653}, 386
  (1999).

\bibitem[Phillips and Cohen(2000)]{PC99}
Phillips, D.~R., and Cohen, T.~D., \emph{Nucl. Phys.}, \textbf{A668}, 45
  (2000).

\bibitem[Park et~al.(2001)]{Pa01}
Park, T.~S., Marcucci, L.~E., Viviani, M., Kievsky, A.,
Rosati, S., Kubodera, K., Min, D.-P., and Rho, M.,
nucl-th/0106025.

\bibitem[Bernard et~al.(1992{\natexlab{a}})]{Br92}
Bernard, V., Kaiser, N., and Mei\ss ner, U.~G., \emph{Nucl. Phys.},
  \textbf{B383}, 442--496 (1992{\natexlab{a}}).

\bibitem[Bernard et~al.(1992{\natexlab{b}})]{Br92B}
Bernard, V., Kaiser, N., Kambor, J., and Mei\ss ner, U.~G., \emph{Nucl. Phys.},
  \textbf{B388}, 315--345 (1992{\natexlab{b}}).

\bibitem[Tonnison et~al.(1998)]{To98}
Tonnison, J., Sandorfi, A.~M., Hoblit, S., and Nathan, A.~M., \emph{Phys. Rev.
  Lett.}, \textbf{80}, 4382--4385 (1998).

\bibitem[Lucas(1994)]{Lu94}
Lucas, M.~A., \emph{Compton scattering from the deuteron at intermediate
  energies}, Ph.D. thesis, University of Illinois (1994), unpublished.

\bibitem[Hornidge et~al.(2000)]{Ho00}
Hornidge, D.~L., et~al., \emph{Phys. Rev. Lett.}, \textbf{84}, 2334--2337
  (2000).

\bibitem[Beane et~al.(1999)]{Be99}
Beane, S.~R., Malheiro, M., Phillips, D.~R., and van Kolck, U., \emph{Nucl.
  Phys.}, \textbf{A656}, 367 (1999).

\bibitem[Wilbois et~al.(1995)]{Wl95}
Wilbois, T., Wilhelm, P., and Arenhovel, H., \emph{Few Bod. Sys.}, \textbf{9},
  263 (1995).

\bibitem[Levchuk and L'vov(1999)]{LL98}
Levchuk, M.~I., and L'vov, A.~I., \emph{Nucl. Phys.},
\textbf{A674}, 449 (2000).

\bibitem[Karakowski and Miller(1999)]{KM99}
Karakowski, J.~J., and Miller, G.~A., \emph{Phys. Rev.}, \textbf{C60}, 014001
  (1999).

\bibitem[Grie\ss hammer and Rupak (2000)]{GR00}
Grie\ss hammer, H.~W., and Rupak, G., nucl-th/0012096.

\bibitem[Schmiedmayer et~al.(1991)]{Sc91}
Schmiedmayer, J., Riehs, P., Harvey, J.~A., and Hill, N.~W.,
\emph{Phys. Rev. Lett.}, \textbf{66}, 1015 (1991).

\bibitem[McGovern (2001)]{McG01}
McGovern, J., \emph{Phys. Rev.}, \textbf{C63}, 064608 (2001), 
and in these proceedings.

\bibitem[Beane et~al.(2001)]{Be02} Beane, S.~R., Malheiro, M.,
McGovern, J., Phillips, D.~R., and van Kolck, U., in preparation.

\bibitem[Bernard et~al.(1996)]{Br96C}
Bernard, V., Kaiser, N., and Mei\ss ner, U.-G., \emph{Z. Phys.}, \textbf{C70},
  483 (1996).

\bibitem[Beane et~al.(1997)]{Be97}
Beane, S.~R., Bernard, V., Lee, H., Mei\ss ner, U.-G., and van Kolck, U.,
  \emph{Nucl. Phys.}, \textbf{A618}, 381 (1997).

\bibitem[Beane (1995)]{Be95}
Beane, S.~R., Lee, C.-Y., van Kolck, U., \emph{Phys. Rev.}, \textbf{C52}, 2914
(1995).

\bibitem[Bergstrom et~al.(1998)]{Be98} Bergstrom, J.~C., Igarashi, R.,
Vogt, J.~M., Kolb, N., Pywell, R.~E., Skopik, D.~M., and Korkmaz, E.,
\emph{Phys. Rev.}, \textbf{C57}, 3203 (1998).

\end{thebibliography}
\end{document}